\documentclass[a4paper,11pt]{article}
\pdfoutput=1 

\usepackage{jheppub} 

\usepackage[T1]{fontenc} 
\usepackage{ifpdf} 
\usepackage{flexisym}
\usepackage{breqn}
\usepackage{comment}

\usepackage{epsf,epsfig,epstopdf}
\usepackage{graphics}
\usepackage{graphicx}
\usepackage[usenames,dvipsnames]{xcolor}
\usepackage{hyperref}
\usepackage{url}
\usepackage{subfig}
\usepackage{amsmath}
\usepackage{multicol}
\usepackage{subfig}
\allowdisplaybreaks
\preprint{TUM-HEP 1454/23}

\title{\boldmath Two-loop master integrals for a planar and a non-planar topology relevant for single top production}

\author[a]{Nikolaos Syrrakos}

\affiliation[a]{Physik Department, Technische Universit\"{a}t M\"{u}nchen, James-Franck-Straße 1, 85748 Garching,
Germany}

\emailAdd{nikolaos.syrrakos@tum.de}

\abstract{We provide analytic results for two-loop four-point master integrals with one massive propagator and one massive leg relevant to single top production. Canonical bases of master integrals are constructed and the Simplified Differential Equations approach is employed for their analytic solution. The necessary boundary terms are computed in closed form in the dimensional regulator, allowing us to obtain analytic results in terms of multiple polylogarithms of arbitrary transcendental weight. We provide explicit solutions of all two-loop master integrals up to transcendental weight six and discuss their numerical evaluation for Euclidean and physical phase-space points.}


\begin{document} 
\maketitle
\flushbottom

\section{Introduction}\label{sec:Introduction}
Providing higher order QCD corrections to the hadronic production of top quarks at the LHC is an important step towards obtaining a more detailed understanding of electroweak symmetry breaking, due to the fact that top quarks receive their mass through interactions with the Higgs background field. Single top production is one of the most prominent top quark production channels at the LHC, with the t-channel process, which involves the production of a single top quark mediated by an exchange of a W boson, accounting for $70\%$ of all single top quarks produced at the LHC~\cite{Giammanco:2017xyn}.

Recently, NNLO corrections to the so-called non-factorisable contributions to the single top t-channel were computed~\cite{Bronnum-Hansen:2021pqc}. The calculation was performed by evaluating all the relevant planar and non-planar two-loop Feynman integrals with one massive leg and up to three massive propagators numerically, using the auxiliary mass flow method~\cite{Liu:2017jxz,Liu:2020kpc,Liu:2021wks,Liu:2022chg}. 

In this paper we take the first step towards the analytic calculation of all two-loop Feynman integrals relevant to non-factorisable contributions to the single top t-channel at NNLO~\cite{Bronnum-Hansen:2021pqc}. We consider a planar and a non-planar topology involving one massive leg and one massive propagator. Our analytic computation is performed in dimensional regularisation, in $d=4-2\epsilon$ space-time dimensions, and relies on Integration-By-Part (IBP) identities~\cite{Chetyrkin:1981qh,Laporta:2000dsw} in order to express all relevant two-loop Feynman integrals to a finite set of so-called master integrals.

\begin{figure}[t]
    \centering
    \begin{tabular}{cc}
    \subfloat[]{\includegraphics[width=7cm]{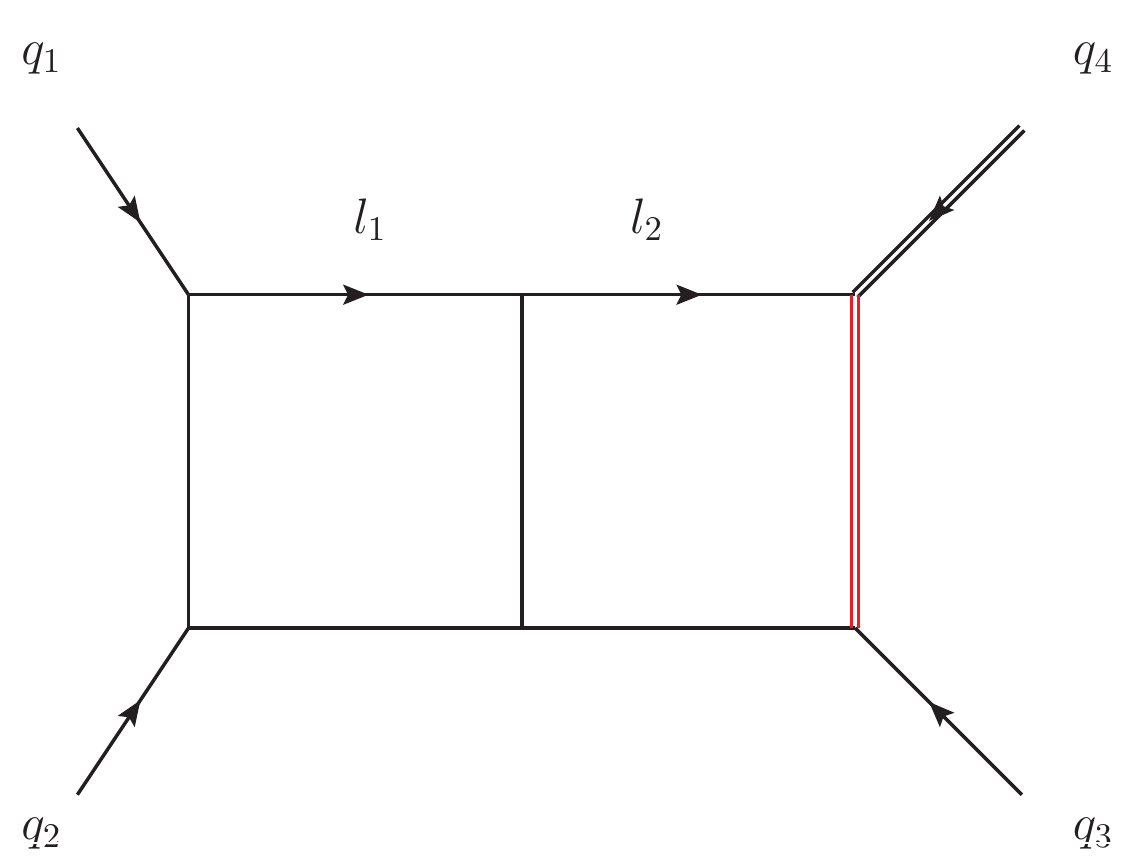}} &
    \subfloat[]{\includegraphics[width=8cm]{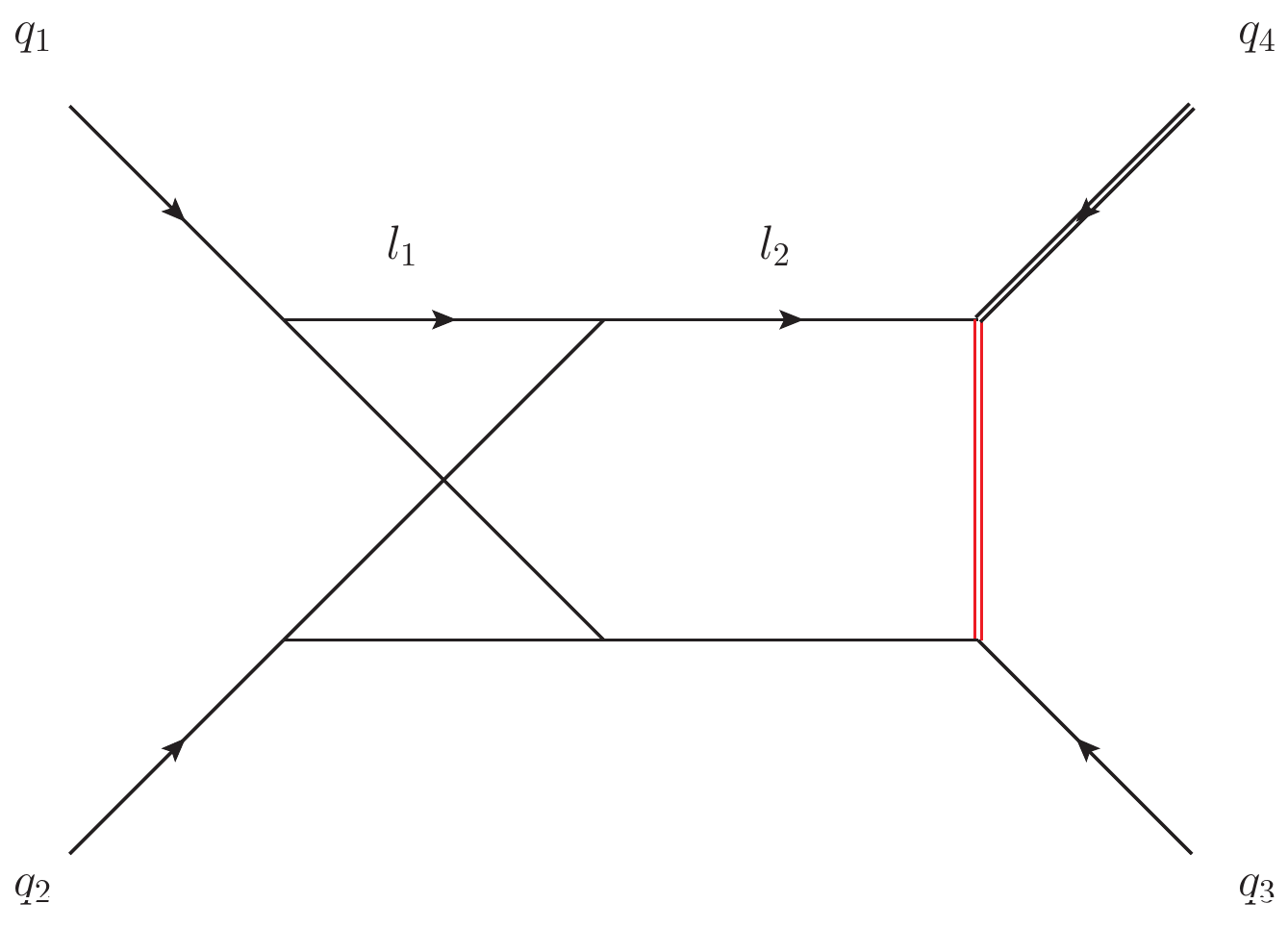}} 
    \end{tabular}
    \caption{Representative diagrams of the top sectors for family $\textbf{PL1}$ (a) and $\textbf{NPL1}$ (b). The black external double line represents the massive top quark, whereas the internal red double line represents the mediated W boson. The diagrams have been drawn using \textsc{JaxoDraw}~\cite{Binosi:2008ig}.}
    \label{fig:topsectors}
\end{figure}

In~\cite{Kotikov:1990kg,Kotikov:1991hm,Kotikov:1991pm,Gehrmann:1999as} it was established that master integrals satisfy differential equations in the external invariants which can be used to evaluate the master integrals themselves. In~\cite{Henn:2013pwa} it was observed that integrals whose singularity structure is only of the logarithmic type and exhibit uniform transcendentality~\cite{Arkani-Hamed:2010pyv}, satisfy a particularly simple system of differential equations, in the so-called canonical form. To that end, we constructed bases of canonical master integrals and derived differential equations for them. We employed a variant of the standard differential equations method, known as Simplified Differential Equations approach (SDE)~\cite{Papadopoulos:2014lla}. The distinction of this approach is that instead of differentiating with respect to the external invariants, one parametrises the external momenta in terms of a dimensionless parameter and then derives differential equations with respect to that parameter only. The master integrals are then determined by analytically solving the canonical differential equations in terms of a well-studied class of special functions, known as multiple polylogarithms (MPLs)~\cite{Goncharov:1998kja,Duhr:2011zq,Duhr:2014woa,Vollinga:2004sn}. 

In order to determine the boundary terms for the solution of the differential equations, we used the method of expansion-by-regions~\cite{Jantzen:2012mw}, which allows us to obtain information about asymptotic limits of individual Feynman integrals. Our analysis allowed us to compute all boundary terms in closed form in the dimensional regulator, allowing us to iteratively solve the differential equations to arbitrary order in $\epsilon$. To exemplify this, we provide explicit results for all two-loop canonical master integrals in terms of MPLs of transcendental weight six. 

Apart from their relevance to the single top production corrections, our results add to the growing list of two-loop four-point Feynman integrals involving internal masses which admit polylogarithmic solutions, see for example~\cite{Wang:2022enl,Long:2021vse,Chen:2021gjv,Becchetti:2019tjy,Bonetti:2020hqh,Bonetti:2022lrk,Heller:2019gkq,Heller:2021gun,Hasan:2020vwn,DiVita:2018nnh,Mastrolia:2017pfy}. The study of these integrals can contribute to the construction of analytic solutions for fast and efficient numerical evaluation in physical regions of phase-space~\cite{Heller:2019gkq,Heller:2021gun}. 

Our paper is structured as follows: in section \ref{sec:Families} we define the integral families under consideration and the relevant kinematics, in section \ref{sec:DE} we describe the construction of canonical bases of master integrals and derive differential equations for them, in section \ref{sec:Boundaries} we compute the necessary boundary terms for the solution of the canonical differential equations and in section \ref{sec:Results} we discuss the analytic solutions of the master integrals, we comment on the complexity of the resulting expressions and report on several numerical checks we performed for Euclidean and physical phase-space points. We give our concluding remarks in section \ref{sec:Conclusion}, and in appendix 
\ref{sec:ApA} we present explicit analytic results for the most complicated canonical master integrals up to $\mathcal{O}(\epsilon^2)$.

\section{Integral families and kinematics}\label{sec:Families}
We consider the analytic calculation of two-loop master integrals for a planar and a non-planar topology involving one massive propagator and one massive leg. The kinematics satisfied by the external momenta are
\begin{equation}
 \sum_{i=1}^4 q_i = 0,\quad q_i^2=0,\, i=1,2,3,\quad q_4^2=m_t^2   
\end{equation}
where $m_t$ denotes the mass of the top quark. The master integrals will depend on the Mandelstam variables $s_{ij}=(q_i+q_j)^2$, with $s_{12}+s_{23}+s_{13}=m_t^2$, as well as the mass of the mediated W boson, $m_w$.

The two-loop integral families are defined as follows:
\begin{equation}
        \mathcal{I}_{topo}(a_1,\ldots ,a_9)=\int \frac{\mathrm{d}^dl_1}{i\pi^{d/2}}\frac{\mathrm{d}^dl_2}{i\pi^{d/2}}\, \frac{\mathrm{e}^{2\gamma_E \epsilon}}{\prod_{i=1}^9 \mathcal{D}_i^{a_i}}
\end{equation}
with $topo \in \{\textbf{PL1},\,\textbf{NPL1}\}$ and the propagators for each family are given in Table \ref{tab:auxtopo}. Throughout this paper we use dimensional regularisation with $d=4-2\epsilon$. In Figure \ref{fig:topsectors} we show two-loop diagrams which are representative of the top sector for each integral family. It is understood that for the purposes of this calculation $a_i>0$ for $i\in\{1,2,3,5,6,8,9\}$ and $a_i<0$ for $i\in\{4,7\}$ for topology $\textbf{PL1}$, whereas $a_i>0$ for $i\in\{1,2,3,5,7,8,9\}$ and $a_i<0$ for $i\in\{4,6\}$ for topology $\textbf{NPL1}$.
\begin{table}[t]
  \centering
  \begin{tabular}{c c c}
    \hline
    \hline
    \multicolumn{1}{c}{Denominator} &
    \multicolumn{1}{c}{Family $\textbf{PL1}$} &
    \multicolumn{1}{c}{Family $\textbf{NPL1}$} \\
    \hline
    \hline
    $D_1$ & $l_1^2$                       & $l_1^2$       \\
    $D_2$ & $l_2^2$                       & $l_2^2$   \\
    $D_3$ & $(l_1 - q_1)^2$               & $(l_1 - q_1)^2$  \\
    $D_4$ & $(l_2 - q_1)^2$               & $(l_2 - q_1)^2$  \\ 
    $D_5$ & $(l_1 - q_1-q_2)^2$           & $(l_2-q_1-q_2)^2$  \\
    $D_6$ & $(l_2-q_1-q_2)^2$             & $(l_1-q_1-q_2-q_3)^2$ \\
    $D_7$ & $(l_1-q_1-q_2-q_3)^2$         & $(l_2-q_1-q_2-q_3)^2-m_w^2$ \\
    $D_8$ & $(l_2-q_1-q_2-q_3)^2-m_w^2$   & $(l_1-l_2+q_2)^2$ \\
    $D_9$ & $(l_1-l_2)^2$                 & $(l_1-l_2)^2$ \\
    \hline
  \end{tabular}
  \caption{Definition of the propagators for the two integral families under consideration. We indicate with $m_w$ the mass of the W boson.}\label{tab:auxtopo}
\end{table}

Using \textsc{Kira}~\cite{Klappert:2020nbg}, a program which implements and automates the Laporta algorithm~\cite{Laporta:2000dsw} for IBPs, we are able to identify 31 and 35 master integrals for families $\textbf{PL1}$ and $\textbf{NPL1}$ respectively. 

For the calculation of the integral families under consideration we employ the SDE approach~\cite{Papadopoulos:2014lla}. This approach relies on the well-established method of deriving differential equations for master integrals of a specific integral family and then using IBP identities to express the result of the differentiation in terms of the same basis of master integrals. The distinction of the SDE approach is that instead of deriving differential equations with respect to the Mandelstam variables $s_{ij}$ and the masses involved, one parameterises the external momenta in terms of a dimensionless parameter, $x$, and then derives differential equations with respect to only that parameter. 

For this particular case, we choose the following parametrisation
\begin{equation}\label{eq:xparam}
    q_1 = -x p_2 ,\quad q_2 = -x p_1,\quad q_3 = -p_1-p_2-p_3,\quad q_4 = (p_1 +p_2) x+p_1+p_2+p_3
\end{equation}
with the $p_i$ momenta satisfying
\begin{equation}
 \sum_{i=1}^4 p_i = 0,\quad p_i^2=0,\, i=1,2,3,4.   
\end{equation}
We are then able to define new Mandelstam variables for the $p_i$ momenta, $S_{ij}=(p_i+p_j)^2$. The definition \eqref{eq:xparam} introduces a mapping between $\{s_{ij},\,m_t\}$ and $\{S_{ij},\,x\}$,
\begin{equation}\label{eq:mapping}
    m_t^2 = S_{12} x (x+1),\quad s_{12} = S_{12} x^2,\quad s_{23} = -x S_{23}.
\end{equation}
The propagators defined in Table \ref{tab:auxtopo} now depend explicitly on $x$ through \eqref{eq:xparam}. We also perform the additional transformation for the loop momenta
\begin{align}
    \textbf{PL1}:& \quad l_1\to k_1 - x p_1 - x p_2,\, l_2\to -k_2 - x p_1 - x p_2, \label{eq:lparam1} \\
    \textbf{NPL1}:& \quad l_1\to k_1 - x p_2,\, l_2\to -k_2 - x p_1 - x p_2. \label{eq:lparam2}
\end{align}
The explicit form of the propagators after the above re-definitions are shown in Table \ref{tab:auxtopox}.
\begin{table}[t]
  \centering
  \begin{tabular}{c c c}
    \hline
    \hline
    \multicolumn{1}{c}{Denominator} &
    \multicolumn{1}{c}{Family $\textbf{PL1}$} &
    \multicolumn{1}{c}{Family $\textbf{NPL1}$} \\
    \hline
    \hline
    $D_1$ & $(k_1-x p_1-x p_2)^2$         & $(k_1-x p_2)^2$       \\
    $D_2$ & $(k_2+x p_1+x p_2)^2$         & $(k_2+x p_1+x p_2)^2$   \\
    $D_3$ & $(k_1 - x p_1)^2$             & $k_1^2$  \\
    $D_4$ & $(k_2 + x p_1)^2$             & $(k_2 + x p_1)^2$  \\ 
    $D_5$ & $k_1^2$                       & $k_2^2$  \\
    $D_6$ & $k_2^2$                       & $(k_1+p_1+p_2+p_3+x p_1)^2$ \\
    $D_7$ & $(k_1+p_1+p_2+p_3)^2$         & $(k_2-p_1-p_2-p_3)^2-m_w^2$ \\
    $D_8$ & $(k_2-p_1-p_2-p_3)^2-m_w^2$   & $(k_1+k_2)^2$ \\
    $D_9$ & $(k_1+k_2)^2$                 & $(k_1+k_2+x p_1)^2$ \\
    \hline
  \end{tabular}
  \caption{Definition of the propagators for the two integral families following \eqref{eq:xparam}, \eqref{eq:lparam1}, \eqref{eq:lparam2}.}\label{tab:auxtopox}
\end{table}

We note that the rescalings of the loop momenta \eqref{eq:lparam1}, \eqref{eq:lparam2} are not necessary for deriving differential equations in the $x$ variable, the parametrisation introduced in \eqref{eq:xparam} is sufficient for that. However, we observe that because of \eqref{eq:lparam1} and \eqref{eq:lparam2}, setting $x=0$ in the propagators defined in Table \ref{tab:auxtopox} results in all denominators appearing in the top-sector scalar integrals, i.e. $\{\mathcal{D}_1,\mathcal{D}_2,\mathcal{D}_3,\mathcal{D}_5,\mathcal{D}_6,\mathcal{D}_8,\mathcal{D}_9\}$ for $\textbf{PL1}$ and $\{\mathcal{D}_1,\mathcal{D}_2,\mathcal{D}_3,\mathcal{D}_5,\mathcal{D}_7,\mathcal{D}_8,\mathcal{D}_9\}$ for $\textbf{NPL1}$, to revert back to $\{\mathcal{D}_5,\mathcal{D}_6,\mathcal{D}_8,\mathcal{D}_9\}$ for $\textbf{PL1}$ and $\{\mathcal{D}_3,\mathcal{D}_5,\mathcal{D}_7,\mathcal{D}_8,\mathcal{D}_9\}$ for $\textbf{NPL1}$. This will become important in section \ref{sec:Boundaries}, where we will compute boundary terms in the limit $x\to 0$, since it will allow us to express several integrals at this limit in terms of two-point functions which are known in closed form using just IBP identities.

We conclude this section by stating that we will work in the Euclidean region defined through the $\{s_{ij},\,m_t,\,m_w\}$ and $\{S_{ij},\,x,\,m_w\}$ variables as
\begin{align}\label{eq:euclidean_region}
   \{s_{ij},\,m_t,\,m_w\}:&\quad s_{12} < 0,\quad s_{23} \leq 0,\quad m_t^2 \leq s_{12} + s_{23},\quad m_w^2 > 0,\\
    \{S_{ij},\,x,\,m_w\}:&\quad x > 0,\quad S_{12} < 0,\quad 0 \leq S_{23} \leq -S_{12},\quad m_w^2 > 0,
\end{align}
where all master integrals are real.

\section{Differential equations}\label{sec:DE}
The method of differential equations~\cite{Kotikov:1990kg,Kotikov:1991hm,Kotikov:1991pm,Gehrmann:1999as} has been established as the currently most powerful approach for the calculation of master integrals, either analytically in terms of special functions~\cite{Goncharov:1998kja,Abreu:2022mfk}, or numerically using power series expansions~\cite{Liu:2017jxz,Liu:2020kpc,Liu:2021wks,Liu:2022chg,Moriello:2019yhu,Hidding:2020ytt,Armadillo:2022ugh}. The observation in~\cite{Henn:2013pwa} that a particular choice of master integrals which exhibit logarithmic singularity structure and are of uniform transcendental weight, satisfy a simple form of differential equations, in the so-called canonical or dlog-form, has lead to numerous results for highly non-trivial cases.

In order to exploit this simplification we constructed bases of canonical master integrals using firstly the \textsc{Mathematica} package \textsc{DLogBasis}~\cite{Henn:2020lye}, which looks in an automated way for integrals with unit leading singularities in $d=4$ spacetime dimensions, through an analysis of iterated residues. For both integral families considered in this paper, \textsc{DLogBasis} was able to provide canonical master integrals for sectors with five denominators and higher. However, there were a few cases of sectors involving more master integrals than the ones \textsc{DLogBasis} was able to provide. For these particular cases we made use of the \textsc{Mathematica} package \textsc{Initial}~\cite{Dlapa:2020cwj} which performs operations on the differential equations to bring them into a canonical form. \textsc{Initial} requires a canonical integral as a first entry and we found it sufficient to apply it only at the level of the specific sectors that we wanted to fix, i.e. at their corresponding maximal cut~\cite{Primo:2016ebd}, and not use it for the differential equations of the full basis of master integrals. For sectors with up to four denominators, we performed a loop-by-loop analysis of the leading singularities for the corresponding integrals in $d=4$ spacetime dimensions, using one-loop results as building blocks. For specific sectors with more than one master integral we used \textsc{Initial}, again at sector level, to obtain the full list of canonical master integrals.

Having the propagators in terms of $x$ as in Table \ref{tab:auxtopox}, the mapping of the kinematic variables \eqref{eq:mapping} and the canonical bases at hand, allows us to derive canonical differential equations for families $\textbf{PL1}$ and $\textbf{NPL1}$. This is achieved by first differentiating the canonical master integrals with respect to $x$ and writing the result in terms of Feynman integrals of the corresponding family, using an in-house \textsc{Mathematica} code. The resulting integrals are then passed through \textsc{Kira} for the IBP reduction. Putting everything together yields the following canonical differential equation
\begin{equation}\label{eq:cansde}
\partial_{x} \textbf{g}=\epsilon \left( \sum_{i=1} \frac{\textbf{M}_i}{x-l_i} \right) \textbf{g}
\end{equation}
where $\textbf{g}$ represents the canonical basis for each family. The residue matrices $\textbf{M}_i$ involve only rational numbers and the poles $l_i$ are functions of the kinematic variables $S_{ij}$. In total we find the following thirteen $l_i$ functions
{\small
\begin{align}\label{eq:poles}
    \bigg\{&-1,0,\frac{m_w^2}{S_{12}},-\frac{r_1+S_{12}}{2 S_{12}},\frac{r_1-S_{12}}{2
   S_{12}},-\frac{S_{12}+S_{23}}{S_{12}},-\frac{m_w^2}{S_{23}},\frac{S_{23}}{S_{12}},-\frac{S_{12} m_w^2+r_2+S_{12} S_{23}}{2 S_{12} S_{23}}, \nonumber \\
   &\frac{-S_{12}
   m_w^2+r_2-S_{12} S_{23}}{2 S_{12} S_{23}},\frac{m_w^2}{S_{12}+S_{23}},\frac{S_{12} m_w^2+r_3-S_{12}^2-S_{12} S_{23}}{2 S_{12}
   \left(S_{12}+S_{23}\right)},-\frac{-S_{12} m_w^2+r_3+S_{12}^2+S_{12} S_{23}}{2 S_{12} \left(S_{12}+S_{23}\right)}\bigg\},
   \end{align}
   }
   where we have defined the square roots
   \begin{align}\label{eq:roots}
       &r_1 = \sqrt{S_{12} \left(4 m_w^2+S_{12}\right)},\quad r_2 = \sqrt{S_{12} \left(4 S_{23}^2 m_w^2+S_{12} \left(m_w^2+S_{23}\right){}^2\right)}, \nonumber \\
       &r_3 = \sqrt{S_{12}
   \left(S_{12} m_w^4+2 \left(S_{12}^2+3 S_{23} S_{12}+2 S_{23}^2\right) m_w^2+S_{12} \left(S_{12}+S_{23}\right){}^2\right)}.
   \end{align}
   Family $\textbf{PL1}$ requires the first ten elements of the singularity vector \eqref{eq:poles}, while family $\textbf{NPL1}$ requires all thirteen of them. Thus, the square root $r_3$ appears only in the differential equation of the non-planar family. Note that the square roots \eqref{eq:roots} do not appear in the definition of the canonical bases, but come from the fact that before partial fractioning, the canonical differential equations have denominators which are polynomials of second order in $x$. 

   The structure of the differential equation \eqref{eq:cansde} allows for an order-by-order in $\epsilon$ integration with respect to $x$, with the result being readily expressed in terms of MPLs~\cite{Goncharov:1998kja}, defined as 
   \begin{align}
        G(a_1,a_2,\ldots, a_n;x) &= \int_0^x \, \frac{\mathrm{dt}}{t-a_1}G(a_2,\ldots, a_n;t),\\
        G(0,\ldots, 0;x) &= \frac{1}{n!}\log^n(x).
    \end{align}
   Indeed, one can write the formal solution of \eqref{eq:cansde} up to $\mathcal{O}(\epsilon^4)$ as follows
   \begin{align}
   \label{eq:solution}
   \textbf{g}&= \epsilon^0 \textbf{b}^{(0)}_{0} + \epsilon \bigg(\sum G_{a}\textbf{M}_{a}\textbf{b}^{(0)}_{0}+\textbf{b}^{(1)}_{0}\bigg) \nonumber \\
   &+ \epsilon^2 \bigg(\sum G_{ab}\textbf{M}_{a}\textbf{M}_{b}\textbf{b}^{(0)}_{0}+\sum G_{a}\textbf{M}_{a}\textbf{b}^{(1)}_{0}+\textbf{b}^{(2)}_{0}\bigg) \nonumber \\
   &+ \epsilon^3 \bigg(\sum G_{abc}\textbf{M}_{a}\textbf{M}_{b}\textbf{M}_{c}\textbf{b}^{(0)}_{0}+\sum G_{ab}\textbf{M}_{a}\textbf{M}_{b}\textbf{b}^{(1)}_{0}+\sum G_{a}\textbf{M}_{a}\textbf{b}^{(2)}_{0}+\textbf{b}^{(3)}_{0}\bigg) \nonumber \\
   &+ \epsilon^4 \bigg(\sum G_{abcd}\textbf{M}_{a}\textbf{M}_{b}\textbf{M}_{c}\textbf{M}_{d}\textbf{b}^{(0)}_{0}+\sum G_{abc}\textbf{M}_{a}\textbf{M}_{b}\textbf{M}_{c}\textbf{b}^{(1)}_{0}\nonumber \\
   &+ \sum G_{ab}\textbf{M}_{a}\textbf{M}_{b}\textbf{b}^{(2)}_{0}+\sum G_{a}\textbf{M}_{a}\textbf{b}^{(3)}_{0}+\textbf{b}^{(4)}_{0}\bigg)
\end{align}
were $ G_{ab\ldots}:=  G(l_a,l_b,\ldots;x)$ represent the MPLs. The $\textbf{b}_0^{(i)}$ terms represent the boundary terms which we compute in the next section, with $i$ indicating the corresponding weight. These results are presented in such a way that each coefficient of $\epsilon^i$ has transcendental weight $i$. If we assign weight $-1$ to $\epsilon$, then \eqref{eq:solution} has uniform weight zero. 

This solution can be trivially extended to higher orders in $\epsilon$, assuming of course one can obtain boundary terms up to the desired order. In fact, as we will show in the next section, for the integral families under consideration, we are able to obtain boundary terms to all orders in $\epsilon$ in closed form, effectively allowing us to extend \eqref{eq:solution} to arbitrary order in $\epsilon$.

\section{Boundary terms}\label{sec:Boundaries}
We consider the limit $x\to 0$ in order to compute the necessary boundary terms $\textbf{b}_0^{(i)}$ in \eqref{eq:solution}~\cite{Canko:2020gqp,Canko:2020ylt}. Our analysis is the same for both integral families. At first we study the solution of the canonical differential equation \eqref{eq:cansde} at the limit $x\to 0$. This solution can be written as
\begin{equation}\label{eq:limzero1}
    \lim_{x\to 0}\textbf{g} = \bigg[\textbf{S} \mathrm{e}^{\epsilon \textbf{D} \log(x)} \textbf{S}^{-1}\bigg]\textbf{b}
\end{equation}
with $\textbf{b}=\sum_{i=0}~ \epsilon^i~ b_0^{(i)}$ the unknown boundary terms, and the matrices $\textbf{S},\, \textbf{D}$ are obtained through the Jordan decomposition of the residue matrix associated to $l_i=0$.

We then consider the canonical basis $\textbf{g}$ and via IBP reduction, we write it in terms of a basis of Feynman integrals $\textbf{I}$, i.e.
\begin{equation}\label{eq:gtoI}
    \textbf{g} = \textbf{T}\textbf{I}.
\end{equation}
Using expansion-by-regions we can obtain the asymptotic limit of these integrals at $x\to 0$. This procedure has been implemented in the code \textsc{asy}~\cite{Jantzen:2012mw}, which is shipped along with \textsc{FIESTA}~\cite{Smirnov:2015mct}. From this analysis we obtain
\begin{equation}\label{eq:expbyregions}
    \lim_{x\to 0}\text{I}_i = \sum\limits_j x^{b_j + a_j \epsilon }\text{I}^{(b_j + a_j \epsilon)}_{i}
\end{equation}
where $a_j$ and $b_j$ are integers and $\text{I}_i$ are the individual members of the basis $\textbf{I}$ of Feynman integrals in \eqref{eq:gtoI}. With $\text{I}^{(b_j + a_j \epsilon)}$ we denote the region-integrals which are given by \textsc{FIESTA} as integrals over Feynman parameters. 

For single-scale integrals \textsc{asy} is not able to provide any output, so in order to have complete control of the right-hand side of \eqref{eq:gtoI} at the limit $x\to 0$, we must supply solutions for these integrals, which are of course known in closed form in $\epsilon$. These are the following,
\begin{align}\label{eq:single_scale}
    &\mathcal{I}_{\textbf{PL1}}(0,0,0,0,1,0,0,2,2) = A(1),\nonumber\\
    &\mathcal{I}_{\textbf{PL1}}(0,1,0,0,2,0,0,0,2) = A(2) \left(-S_{12}\right){}^{-2 \epsilon } x^{-4 \epsilon
   -2},\nonumber\\
   &\mathcal{I}_{\textbf{PL1}}(1,0,0,0,2,0,0,2,0) = A(3) \left(-S_{12}\right){}^{-\epsilon } x^{-2 (\epsilon +1)},\nonumber\\
   &\mathcal{I}_{\textbf{PL1}}(1,0,0,1,1,0,0,0,1) = A(4)
   \left(-S_{12}\right){}^{-2 \epsilon } x^{-4 \epsilon },\nonumber\\
   &\mathcal{I}_{\textbf{PL1}}(1,1,0,0,1,1,0,0,0) = A(5) \left(-S_{12}\right){}^{-2 \epsilon } x^{-4 \epsilon }\\
   &\mathcal{I}_{\textbf{NPL1}}(0,0,1,0,0,0,1,1,0) = B(1),\nonumber\\
   &\mathcal{I}_{\textbf{NPL1}}(0,1,1,0,0,0,0,1,0) = B(2) \left(-S_{12}\right){}^{-2 \epsilon } x^{2-4 \epsilon
   },\nonumber\\
   &\mathcal{I}_{\textbf{NPL1}}(1,1,0,0,1,0,0,1,0) = B(3) \left(-S_{12}\right){}^{-2 \epsilon } x^{-4 \epsilon },\nonumber\\
   &\mathcal{I}_{\textbf{NPL1}}(1,1,1,0,1,0,0,1,1) = B(4)
   \left(-S_{12}\right){}^{-2 \epsilon } x^{-4 (\epsilon +1)}\nonumber
\end{align}
where the $A(i)$ and $B(i)$ are functions independent of $x$,
\begin{align}\label{eq:AB}
    A(1) =& -\frac{e^{2 \gamma  \epsilon } \Gamma (1-\epsilon ) \Gamma (-\epsilon ) \Gamma (\epsilon +1) \Gamma (2 \epsilon +1) \left(m_w^2\right){}^{-2 \epsilon
   -1}}{\Gamma (2-\epsilon )},\nonumber\\
   A(2) =& \frac{e^{2 \gamma  \epsilon } \Gamma (1-\epsilon ) \Gamma (-\epsilon )^2 \Gamma (2 \epsilon +1)}{S_{12} \Gamma (1-3 \epsilon
   )},\nonumber\\
   A(3) =& \frac{e^{2 \gamma  \epsilon } \Gamma (1-\epsilon ) \Gamma (-\epsilon ) \Gamma (\epsilon ) \Gamma (\epsilon +1) \left(m_w^2\right){}^{-\epsilon }}{S_{12}
   \Gamma (1-2 \epsilon )},\nonumber\\
   A(4) =& \frac{e^{2 \gamma  \epsilon } \Gamma (1-2 \epsilon )^2 \Gamma (1-\epsilon )^2 \Gamma (\epsilon ) \Gamma (2 \epsilon )}{\Gamma (2-3
   \epsilon ) \Gamma (2-2 \epsilon )},\nonumber\\
   A(5) =& \frac{e^{2 \gamma  \epsilon } \Gamma (1-\epsilon )^4 \Gamma (\epsilon )^2}{\Gamma (2-2 \epsilon )^2},\\
   B(1) =& -\frac{e^{2
   \gamma  \epsilon } \Gamma (1-\epsilon )^2 \Gamma (\epsilon ) \Gamma (2 \epsilon -1) \left(m_w^2\right){}^{1-2 \epsilon }}{\Gamma (2-\epsilon )},\nonumber\\
   B(2) =& \frac{S_{12}
   e^{2 \gamma  \epsilon } \Gamma (1-\epsilon )^3 \Gamma (2 \epsilon -1)}{\Gamma (3-3 \epsilon )},\nonumber\\
   B(3) =& \frac{e^{2 \gamma  \epsilon } \Gamma (1-2 \epsilon )^2 \Gamma
   (1-\epsilon )^2 \Gamma (\epsilon ) \Gamma (2 \epsilon )}{\Gamma (2-3 \epsilon ) \Gamma (2-2 \epsilon )},\nonumber\\
   B(4) =& -\frac{e^{2 \gamma  \epsilon } \Gamma (1-\epsilon ) \Gamma (2 \epsilon +1)}{2 S_{12}^2 \epsilon ^4} \bigg(-\frac{2 \Gamma (1-2 \epsilon )^4 \Gamma (\epsilon +1) \Gamma (2 \epsilon +1)^2}{\Gamma (1-4 \epsilon )^2 \Gamma (4 \epsilon +1)}\nonumber\\
   &-\frac{8 \epsilon ^2 \Gamma (1-2 \epsilon ) \Gamma (1-\epsilon )}{(\epsilon +1) (2 \epsilon +1) \Gamma (1-4 \epsilon )}\, _3F_2(1,1,2 \epsilon +1;\epsilon +2,2 \epsilon +2;1)\nonumber\\
   &+\frac{\Gamma (1-2 \epsilon ) \Gamma (1-\epsilon ) \Gamma (\epsilon +1)}{\Gamma (1-3 \epsilon )}\, _3F_2(1,-4 \epsilon ,-2 \epsilon ;1-3 \epsilon ,1-2 \epsilon ;1)\nonumber\\
   &-\frac{\Gamma (1-\epsilon )^2}{\Gamma (1-3 \epsilon )}\, _4F_3(1,1-\epsilon ,-4 \epsilon ,-2 \epsilon ;1-3 \epsilon ,1-2 \epsilon ,1-2 \epsilon ;1)\bigg)\nonumber
\end{align}
We can then obtain the limit $x\to 0$ of $\textbf{g}$ using \eqref{eq:gtoI}, the closed-form solutions for the single-scale integrals and \eqref{eq:expbyregions} as follows
\begin{equation}\label{eq:limzero2}
    \lim_{x\to 0}\textbf{g} = \left.\lim _{x \rightarrow 0} \mathbf{T} \mathbf{I}\right|_{\mathcal{O}\left(x^{0+a_{j} \epsilon}\right)}
\end{equation}
where the right-hand side implies that apart from the terms $x^{a_j \epsilon}$ coming from \eqref{eq:expbyregions}, we expand around $x=0$, keeping only terms of order $x^0$.

Equating the right-hand side of \eqref{eq:limzero1} and \eqref{eq:limzero2} allows us to construct the following relation
\begin{equation}\label{eq:boundaries}
    \bigg[\textbf{S} \mathrm{e}^{\epsilon \textbf{D} \log(x)} \textbf{S}^{-1}\bigg]\textbf{b} = \left.\lim _{x \rightarrow 0} \mathbf{T} \mathbf{I}\right|_{\mathcal{O}\left(x^{0+a_{j} \epsilon}\right)}.
\end{equation}
Equation \eqref{eq:boundaries} allows us to fix all boundary terms. First of all we obtain relations between boundary terms of several basis elements
\begin{align}
   \textbf{PL1}:\, \big\{&b_1 = 0,b_3 = -b_2,b_5 = 0,b_6 = -b_2,b_8 = 0,b_9 = 0,b_{10} = 0,b_{11} = 0,b_{12} = 0, \nonumber \\
   &b_{13} = 0,b_{15} = 0,b_{17} = -b_{14},b_{18} =
   -\frac{b_4}{4},b_{19} = 0,b_{20} = 0,b_{21} = -\frac{b_4}{8},b_{23} = 0,\nonumber\\
   &b_{24} = 0,b_{25} = 0,b_{26} = 0,b_{27} = 0,b_{28} = 0,b_{29} = 0,b_{30} =
   b_4-b_{14}+b_{16},\nonumber\\
   &b_{31} = -\frac{3 b_4}{2}+3 b_{14}-b_{16}\big\},\label{eq:b1} \\
   \textbf{NPL1}:\, \big\{&b_1 = 0,b_3 = -b_2,b_5 = 0,b_6 = 0,b_7 = b_2,b_8 = b_2,b_9 = 0,b_{10} = 0,b_{11} = 0,\nonumber\\
   &b_{12} = 0,b_{13} = 0,b_{14} = -b_{16},b_{15} =
   -\frac{b_4}{4},b_{17} = 0,b_{18} = 0,b_{19} = 0,b_{20} = 0,\nonumber\\
   &b_{21} = b_{16},b_{22} = 0,b_{23} = 0,b_{24} = 0,b_{25} = b_{16},b_{26} = 0,b_{27} =
   b_{16},b_{28} = 0,\nonumber\\
   &b_{29} = \frac{b_4}{4},b_{30} = 0,b_{31} = \frac{b_4}{4},b_{33} = \frac{b_4}{4}+b_{32}\big\}\label{eq:b2}.
\end{align}
The above relations leave the following boundary terms to be fixed
\begin{align}
   \textbf{PL1}:\,& \left\{b_2,b_4,b_7,b_{14},b_{16},b_{22}\right\},\\
   \textbf{NPL1}:\,& \left\{b_2,b_4,b_{16},b_{32},b_{34},b_{35}\right\}.
\end{align}
In order to compute the remaining boundary terms, we return to \eqref{eq:limzero2} and observe that several of them are related to single-scale integrals as defined in \eqref{eq:single_scale}, \eqref{eq:AB}. 

For family $\textbf{PL1}$ these are
\begin{align}
    &b_2= -\left(A(1) (\epsilon -1) \epsilon  m_w^2\right),\\
   &b_4=A(2) \epsilon ^2 \left(-S_{12}\right){}^{1-2 \epsilon },\\
   &b_7=A(3) \epsilon ^2 \left(-S_{12}\right){}^{1-\epsilon },\\
  &b_{14}= -\frac{1}{2} A(4) \epsilon ^2 \left(6 \epsilon ^2-5 \epsilon +1\right) \left(-S_{12}\right){}^{-2 \epsilon },\\
   &b_{16}=A(5) (1-2 \epsilon )^2 \epsilon ^2 \left(-S_{12}\right){}^{-2 \epsilon }.
\end{align}

For family $\textbf{NPL1}$ these are
\begin{align}
    &b_4=2 B(2) \epsilon  \left(18 \epsilon ^3-27 \epsilon ^2+13 \epsilon -2\right) \left(-S_{12}\right){}^{-2 \epsilon -1}, \\
   &b_{16}=\frac{1}{2} \epsilon  \left(6 \epsilon ^2-5 \epsilon +1\right) \left(-S_{12}\right){}^{-2 \epsilon -1} \left(B(3) S_{12} \epsilon +B(2) (2-3 \epsilon )\right),\\
   &b_{32}=B(4) \epsilon ^4 \left(-S_{12}\right){}^{2-2 \epsilon }.
\end{align}
So we are left with the following boundary terms to be fixed
\begin{align}
   \textbf{PL1}:\,& \left\{b_{22}\right\},\\
   \textbf{NPL1}:\,& \left\{b_2,b_{34},b_{35}\right\}.
\end{align}
Looking again at \eqref{eq:limzero2}, we find that to completely fix these boundary terms we need to compute the following region-integrals, 8 for $\textbf{PL1}$ and 23 for $\textbf{NPL1}$,
\begin{align}
     \textbf{PL1}:\,\{&\text{I}_{\textbf{PL1}}^0(0,0,1,0,0,0,0,1,1),\text{I}_{\textbf{PL1}}^0(0,0,1,0,0,0,0,2,2),\nonumber\\
     &\text{I}_{\textbf{PL1}}^0(1,0,0,0,0,0,0,1,1),\text{I}_{\textbf{PL1}}^1(1,0,0,0,0,0,0,1,1),\nonumber\\
     &\text{I}_{\textbf{PL1}}^0(1,0,0,0,0,0,0,2,2),\text{I}_{\textbf{PL1}}^1(1,
   0,0,0,0,0,0,2,2),\nonumber\\
   &\text{I}_{\textbf{PL1}}^0(1,0,0,0,1,0,0,1,1),\text{I}_{\textbf{PL1}}^0(1,0,0,0,1,0,0,2,1)\},\label{eq:regions_pl}\\
   \textbf{NPL1}:\,\{&\text{I}_{\textbf{NPL1}}^0(-1,0,1,0,0,0,1,0,1),\text{I}_{\textbf{NPL1}}^0(0,0,1,0,0,0,1,0,1),\nonumber\\
   &\text{I}_{\textbf{NPL1}}^1(-1,0,1,0,0,0,1,0,1),\text{I}_{\textbf{NPL1}}^0(-1,1,1,0,0,0,1,1,0),\nonumber\\
   &\text{I}_{\textbf{NPL1}}^0(-1,1,1,0,1,0,1,0,1),\text{I}_{\textbf{NPL1}}^
   1(0,0,1,0,0,0,1,0,1),\nonumber\\
   &\text{I}_{\textbf{NPL1}}^0(0,1,1,0,0,0,1,0,1),\text{I}_{\textbf{NPL1}}^0(0,1,1,0,0,0,1,1,0),\nonumber\\
   &\text{I}_{\textbf{NPL1}}^0(1,-1,0,0,0,0,1,0,1),\text{I}_{\textbf{NPL1}}^1(1,-1,0,0,0,0,1,0,1),\nonumber\\
   &\text{I}_{\textbf{NPL1}}^0(1,-1,0,0,0,0,
   1,1,0),\text{I}_{\textbf{NPL1}}^1(1,-1,0,0,0,0,1,1,0),\nonumber\\
   &\text{I}_{\textbf{NPL1}}^0(1,-1,1,0,1,0,1,0,1),\text{I}_{\textbf{NPL1}}^0(1,0,0,0,0,0,1,0,1),\nonumber\\
   &\text{I}_{\textbf{NPL1}}^1(1,0,0,0,0,0,1,0,1),\text{I}_{\textbf{NPL1}}^0(1,0,0,0,0,0,1,1,0),\nonumber\\
   &\text{I}_{\textbf{NPL1}}
   ^1(1,0,0,0,0,0,1,1,0),\text{I}_{\textbf{NPL1}}^0(1,0,0,0,1,0,1,0,1),\nonumber\\
   &\text{I}_{\textbf{NPL1}}^0(1,1,-1,0,1,0,1,1,0),\text{I}_{\textbf{NPL1}}^0(1,1,0,0,0,0,1,1,0),\nonumber\\
   &\text{I}_{\textbf{NPL1}}^0(1,1,1,0,-1,0,1,1,0),\text{I}_{\textbf{NPL1}}^0(-1,1,1,0,0,0
   ,1,1,1),\nonumber\\
   &\text{I}_{\textbf{NPL1}}^0(1,-1,0,0,1,0,1,1,1)\}\label{eq:regions_npl}.
\end{align}
The notation used in \eqref{eq:regions_pl}, \eqref{eq:regions_npl} is meant to be interpreted as follows. Consider for example $\text{I}_{\textbf{NPL1}}^0(1,1,1,0,-1,0,1,1,0)$. This region-integral comes for the asymptotic limit of $\mathcal{I}_{\textbf{NPL1}}(1,1,1,0,-1,0,1,1,0)$, i.e.
\begin{equation}\label{eq:examp1}
    \lim_{x\to 0}\mathcal{I}_{\textbf{NPL1}}(1,1,1,0,-1,0,1,1,0) = x^0 \text{I}_{\textbf{NPL1}}^0(1,1,1,0,-1,0,1,1,0) + \mathcal{O}(x^1).
\end{equation}
Note that from \eqref{eq:regions_pl}, \eqref{eq:regions_npl} we see that we only have to compute region-integrals that scale like $x^0$ or $x^1$. 

In principle one can compute all region-integrals one-by-one by doing the integrations over Feynman parameters. We choose to follow two different approaches for their calculation. We first separate them according to their scaling in $x$. Those that scale as $x^1$ involve at most three integrations over Feynman parameters and are easily computed in closed form in $\epsilon$,
\begin{align}
    &\text{I}_{\textbf{PL1}}^1(1,0,0,0,0,0,0,1,1) = \frac{S_{12} e^{2 \gamma  \epsilon } \epsilon  \Gamma (-\epsilon ) \Gamma (2 \epsilon ) \Gamma (\epsilon +1)
   \left(m_w^2\right){}^{-2 \epsilon }}{\epsilon ^2-3 \epsilon +2},\\
   &\text{I}_{\textbf{PL1}}^1(1,0,0,0,0,0,0,2,2) = -\frac{S_{12} e^{2 \gamma  \epsilon } (\epsilon +1)
   \Gamma (-\epsilon ) \Gamma (\epsilon +1) \Gamma (2 \epsilon +2) \left(m_w^2\right){}^{-2 (\epsilon +1)}}{\epsilon ^2-3 \epsilon +2},\\
&\text{I}_{\textbf{NPL1}}^1(-1,0,1,0,0,0,1,0,1) = \frac{\left(-S_{12}-S_{23}\right) e^{2 \gamma  \epsilon } \Gamma (-\epsilon ) \Gamma (\epsilon +1) \Gamma (2
   \epsilon -1) \left(m_w^2\right){}^{1-2 \epsilon }}{\epsilon -2},\\
   &\text{I}_{\textbf{NPL1}}^1(0,0,1,0,0,0,1,0,1) = \frac{S_{23} e^{2 \gamma  \epsilon } \Gamma (1-\epsilon
   )^2 \Gamma (2 \epsilon ) \Gamma (\epsilon +1) \left(m_w^2\right){}^{-2 \epsilon }}{\Gamma (3-\epsilon )},\\
   &\text{I}_{\textbf{NPL1}}^1(1,-1,0,0,0,0,1,0,1) = \frac{S_{12}
   e^{2 \gamma  \epsilon } (\epsilon -1) \Gamma (3-2 \epsilon ) \Gamma (1-\epsilon )^2 \Gamma (\epsilon ) \Gamma (2 \epsilon -1) \left(m_w^2\right){}^{1-2 \epsilon
   }}{\Gamma (2-2 \epsilon ) \Gamma (3-\epsilon )},\\
   &\text{I}_{\textbf{NPL1}}^1(1,-1,0,0,0,0,1,1,0) = \frac{\pi ^2 e^{2 \gamma  \epsilon } \left(S_{12} (\epsilon -1)+S_{23} \epsilon \right) \csc ^2(\pi  \epsilon ) \sec (\pi  \epsilon ) \left(m_w^2\right){}^{1-2 \epsilon
   }}{(\epsilon -2) \Gamma (2-2 \epsilon )},\\
   &\text{I}_{\textbf{NPL1}}^1(1,0,0,0,0,0,1,0,1) = -\frac{S_{12} e^{2 \gamma  \epsilon } \Gamma (1-\epsilon )^2 \Gamma (2 \epsilon )
   \Gamma (\epsilon +1) \left(m_w^2\right){}^{-2 \epsilon }}{\Gamma (3-\epsilon )},\\
   &\text{I}_{\textbf{NPL1}}^1(1,0,0,0,0,0,1,1,0) = \frac{\left(-S_{12}-S_{23}\right) e^{2
   \gamma  \epsilon } \Gamma (1-\epsilon )^2 \Gamma (2 \epsilon ) \Gamma (\epsilon +1) \left(m_w^2\right){}^{-2 \epsilon }}{\Gamma (3-\epsilon )}.
\end{align}

For the region-integrals that scale as $x^0$ we consider first the original integral and study its asymptotic limit in momentum space~\cite{Smirnov:2012gma}. In all relevant cases we find that this analysis allows us to write all region-integrals that scale as $x^0$, as single-scale integrals without the need to perform any integrations, but by employing IBP identities. Below we give an explicit example.

Consider the region-integral $\text{I}_{\textbf{NPL1}}^0(1,1,1,0,-1,0,1,1,0)$. As we showed in \eqref{eq:examp1}, this appears in the asymptotic limit of $\mathcal{I}_{\textbf{NPL1}}(1,1,1,0,-1,0,1,1,0)$ for $x\to 0$. We therefore look at $\mathcal{I}_{\textbf{NPL1}}(1,1,1,0,-1,0,1,1,0)$ in momentum space:
\begin{align}\label{eq:examp2}
    \mathcal{I}_{\textbf{NPL1}}(1,1,1,0,-1,0,1,1,0) = \int \frac{\mathrm{d}^dk_1}{i\pi^{d/2}}\frac{\mathrm{d}^dk_2}{i\pi^{d/2}}&\, \frac{\mathrm{e}^{2\gamma_E \epsilon}}{(k_1-x p_1)^2(k_2-x p_1-x p_2)^2k_1^2}\nonumber\\
    \times &\frac{k_2^2}{((k_2-p_1-p_2-p_3)^2-m_w^2)(k_1+k_2)^2}.
\end{align}
Applying expansion-by-regions in momentum space amounts to performing a Taylor expansion of the integrand~\cite{Smirnov:2012gma} in terms of $x$. Keeping terms up to order $\mathcal{O}(x^0)$ effectively means setting $x=0$ in \eqref{eq:examp2}. This yields
\begin{equation}
    \int \frac{\mathrm{d}^dk_1}{i\pi^{d/2}}\frac{\mathrm{d}^dk_2}{i\pi^{d/2}}\, \frac{\mathrm{e}^{2\gamma_E \epsilon}}{(k_1^2)^2((k_2-p_1-p_2-p_3)^2-m_w^2)(k_1+k_2)^2}
\end{equation}
The above integral can be associated to the $\textbf{NPL1}$ family as $\mathcal{I}_{\textbf{NPL1}}(0,0,2,0,0,0,1,1,0)$. Using IBP identities we can rewrite this integral in terms of the single-scale integrals defined in \eqref{eq:single_scale}, \eqref{eq:AB}. Thus we have
\begin{align}
    \text{I}_{\textbf{NPL1}}^0(1,1,1,0,-1,0,1,1,0) =&\,
    \mathcal{I}_{\textbf{NPL1}}(0,0,2,0,0,0,1,1,0)\nonumber\\
     =& \frac{2 \epsilon -1}{m_w^2}\mathcal{I}_{\textbf{NPL1}}(0,0,1,0,0,0,1,1,0).
\end{align}
In the same way we can relate all remaing region-integrals using expansion-by-regions in momentum space and expressing the result in terms of the known single-scale integrals of $\mathcal{I}_{\textbf{PL1}}(0,0,0,0,1,0,0,2,2)$ and  $\mathcal{I}_{\textbf{NPL1}}(0,0,1,0,0,0,1,1,0)$ through IBP identities. The results read
\begin{align}
    &\text{I}_{\textbf{PL1}}^0(0,0,1,0,0,0,0,1,1) = -\frac{A(1) m_w^4}{(4-2 \epsilon )^2-7 (4-2 \epsilon )+12},\\
    &\text{I}_{\textbf{PL1}}^0(0,0,1,0,0,0,0,2,2) =
   A(1),\\
   &\text{I}_{\textbf{PL1}}^0(1,0,0,0,0,0,0,1,1) = -\frac{A(1) m_w^4}{(4-2 \epsilon )^2-7 (4-2 \epsilon )+12},\\
   &\text{I}_{\textbf{PL1}}^0(1,0,0,0,0,0,0,2,2) =
   A(1),\\
   &\text{I}_{\textbf{PL1}}^0(1,0,0,0,1,0,0,1,1) = -\frac{A(1) m_w^2}{2 \epsilon },\\
   &\text{I}_{\textbf{PL1}}^0(1,0,0,0,1,0,0,2,1) = A(1),\\
   &\text{I}_{\textbf{NPL1}}^0(-1,0,1,0,0,0,1,0,1) = 0,\\
   &\text{I}_{\textbf{NPL1}}^0(-1,1,1,0,0,0,1,1,0) = 0,\\
   &\text{I}_{\textbf{NPL1}}^0(-1,1,1,0,0,0,1,1,1) =
   0,\\
   &\text{I}_{\textbf{NPL1}}^0(-1,1,1,0,1,0,1,0,1) = 0,\\
   &\text{I}_{\textbf{NPL1}}^0(0,0,1,0,0,0,1,0,1) = B(1),\\
   &\text{I}_{\textbf{NPL1}}^0(0,1,1,0,0,0,1,0,1) =
   \frac{B(1)}{m_w^2},\\
   &\text{I}_{\textbf{NPL1}}^0(0,1,1,0,0,0,1,1,0) = \frac{B(1)}{m_w^2},\\
   &\text{I}_{\textbf{NPL1}}^0(1,-1,0,0,0,0,1,0,1) = B(1)
   m_w^2,\\
   &\text{I}_{\textbf{NPL1}}^0(1,-1,0,0,0,0,1,1,0) = B(1) m_w^2,\\
   &\text{I}_{\textbf{NPL1}}^0(1,-1,0,0,1,0,1,1,1) = \frac{B(1) (2 \epsilon
   -1)}{m_w^2},\\
   &\text{I}_{\textbf{NPL1}}^0(1,-1,1,0,1,0,1,0,1) = \frac{B(1) (2 \epsilon -1)}{m_w^2},\\
   &\text{I}_{\textbf{NPL1}}^0(1,0,0,0,0,0,1,0,1) =
   B(1),\\
   &\text{I}_{\textbf{NPL1}}^0(1,0,0,0,0,0,1,1,0) = B(1),\\
   &\text{I}_{\textbf{NPL1}}^0(1,0,0,0,1,0,1,0,1) =
   \frac{B(1)}{m_w^2},\\
   &\text{I}_{\textbf{NPL1}}^0(1,1,-1,0,1,0,1,1,0) = 0,\\
   &\text{I}_{\textbf{NPL1}}^0(1,1,0,0,0,0,1,1,0) =
   \frac{B(1)}{m_w^2},\\
   &\text{I}_{\textbf{NPL1}}^0(1,1,1,0,-1,0,1,1,0) = \frac{B(1) (2 \epsilon -1)}{m_w^2}.
\end{align}

The above analysis allows us to fix all remaining boundary terms. We are able therefore to obtain all boundary terms in closed form in $\epsilon$. At most these include the generalised hypergeometric functions
\begin{align*}
    &\, _3F_2(1,1,2 \epsilon +1;\epsilon +2,2 \epsilon +2;1)\\
    &\, _3F_2(1,-4 \epsilon ,-2 \epsilon ;1-3 \epsilon ,1-2 \epsilon ;1)\\
    &\, _4F_3(1,1-\epsilon ,-4 \epsilon ,-2 \epsilon ;1-3 \epsilon ,1-2 \epsilon ,1-2 \epsilon ;1)
\end{align*}
which can be expanded to arbitrary order in $\epsilon$ using the \textsc{Mathematica} package \textsc{HypExp}~\cite{Huber:2007dx}.

\section{Results and checks}\label{sec:Results}
As we stated in the introduction, the integral families considered in this paper are relevant for NNLO corrections to single top production via the mediation of a $W$ boson~\cite{Bronnum-Hansen:2021pqc}. Thus, it is sufficient to compute the corresponding master integrals up to transcendental weight four, or due to our chosen normalisation of the canonical master integrals, up to $\mathcal{O}(\epsilon^4)$. However the structure of the differential equation \eqref{eq:cansde} and the closed-form solutions of the boundary terms we obtained in the previous section, allow us to obtain analytic solutions for all two-loop master integrals in terms of MPLs of arbitrary weight.

In practise we see that going beyond order $\mathcal{O}(\epsilon^4)$ results in a rapid increase in the complexity of the analytic expressions due to the emergence of a huge number of MPLs with weight five and six. For the integration of the differential equation \eqref{eq:cansde} in terms of MPLs we relied on the \textsc{Mathematica} package \textsc{PolyLogTools}~\cite{Duhr:2019tlz}. In Table \ref{tab:mpls} we show an analysis of the number of MPLs for each weight up to six. 
\begin{table}[h!]
\begin{center}
 \begin{tabular}{| c | c | c | c | c | c | c | c |} 
 \hline
 \textbf{Family} & \textbf{$W=1$} & \textbf{$W=2$} & \textbf{$W=3$} & \textbf{$W=4$} & \textbf{$W=5$} & \textbf{$W=6$} & \textbf{Total} \\
 \hline
 $\textbf{PL1}$ & 5 & 38 & 305 & 2346 & 16865 & 71436 & 90995\\ 
 \hline
 $\textbf{NPL1}$ & 6 & 58 & 541 & 4556 & 35064 & 156150 & 196375\\  
 \hline
\end{tabular}
\end{center}
\caption{Number of MPLs per transcendental weight.}
\label{tab:mpls}
\end{table}

Our results can be evaluated numerically in any region of phase-space, both the Euclidean defined in \eqref{eq:euclidean_region} and the physical region
\begin{align}\label{eq:physical_region}
    \{s_{ij},\,m_t,\,m_w\}:&\quad s_{12}>m_t^2>m_w^2>0>s_{13},s_{23},\\
    \{S_{ij},\,x,\,m_w\}:&\quad x<-1,\quad S_{12}>0,\quad S_{23}<0,\quad m_w^2>0.
\end{align}
For the physical region one needs to give a small imaginary part to $s_{12},\, s_{13},\, s_{23}$. For the physical region \eqref{eq:physical_region} the correct result is produced by ensuring that $s_{12}$ and $m_t^2=s_{12}+s_{23}+s_{13}$ have a small positive imaginary part. By studying the single-scale integrals that appear in our canonical bases, we can determine the sign of the small imaginary part for the $\{S_{ij},\,x\}$ variables~\cite{Canko:2020ylt}.
\begin{align}
    (-s_{12})^{(-2\epsilon)} &= (-S_{12})^{(-2\epsilon)} (x^2)^{(-2\epsilon)},\\
    (-m_t^2)^{(-2\epsilon)} &= (-S_{12})^{(-2\epsilon)} (x)^{(-2\epsilon)}(1+x)^{(-2\epsilon)}.
\end{align}
We find that the correct result in the physical region \eqref{eq:physical_region} is produced by giving a small negative imaginary part to $x$ and $S_{12}$.

We performed several checks of our analytic results up to weight four for both Euclidean and physical phase-space points against \textsc{pySecDec}~\cite{Borowka:2017idc}, a program that automates the sector decomposition algorithm for the numerical computation of multiloop Feynman integrals~\cite{Heinrich:2008si}, and against \textsc{AMFLow}~\cite{Liu:2022chg}, which is a program that evaluates Feynman integrals numerically based one the auxiliary mass flow method~\cite{Liu:2017jxz,Liu:2020kpc,Liu:2021wks}. All MPLs were evaluated numerically using \textsc{PolyLogTools} and \textsc{GiNaC}~\cite{Vollinga:2004sn}. The weight five and six expressions can in principle also be evaluated, but their complexity makes their numerical evaluation highly inefficient. We checked our solutions for all master integrals for two Euclidean points. We also checked all master integrals for two physical phase-space points. For master integrals with seven denominators we were unable to obtain any numerical results from \textsc{pySecDec} for physical phase-space points, but we were able to check our solutions for these particular sectors against \textsc{AMFlow}. We report excellent agreement for all performed checks. In Table \ref{tab:points} we show the chosen kinematic points in the $\{S_{ij},x,\,m_w\}$ variables.
\begin{table}[h!]
    \centering
    \begin{tabular}{c}
       \textbf{Euclidean points}  \\
       $\left\{x\to \frac{45}{11},S_{12}\to -\frac{968}{45},S_{23}\to \frac{704}{45},m_w^2\to 974\right\}$  \\
       \hfill \\
       $\left\{x\to \frac{31}{5},S_{12}\to -\frac{700}{31},S_{23}\to \frac{215}{31},m_w^2\to 974\right\}$\\
       \hline
       \hline
       \textbf{Physical points} \\
       $\left\{x\to -\frac{104337}{74408},S_{12}\to \frac{5536550464}{104337},S_{23}\to -\frac{385359032}{104337},m_w^2\to 6400\right\}$\\
        \hfill \\
       $\left\{x\to -\frac{51824}{21895},S_{12}\to \frac{479391025}{51824},S_{23}\to -\frac{87908425}{12956},m_w^2\to 6400\right\}$
    \end{tabular}
    \caption{Kinematic points that were used for numerical checks.}
    \label{tab:points}
\end{table}
In \eqref{eq:num1}-\eqref{eq:num4} we give numerical results for the most complicated of the top-sector canonical master integrals of each integral family for the first Euclidean and the first physical point defined in Table \ref{tab:points}. The complexity of the expressions is gauged according to the total number of MPLs they contain. For family $\textbf{PL1}$ the most complicated canonical master integral is $g_{30}$ containing 1166 MPLs and for family $\textbf{NPL1}$ the most complicated canonical master integral is $g_{34}$ containing 2149 MPLs.
\begin{align}\label{eq:num1}
    g_{30,\, E_1}^{\textbf{PL1}} =& \frac{1}{4} - 2.81748762244805114730354528779789763 \epsilon \nonumber\\
    &+ 15.95471394152217759709401049935433650 \epsilon^2\nonumber\\
    &-57.4690121428642202790795370777756439 \epsilon^3 \nonumber \\
    &+143.3537514494667074193691678448309506 \epsilon^4
\end{align}
\begin{align}\label{eq:num2}
    g_{30,\, P_1}^{\textbf{PL1}} =& \frac{1}{4} - 5.7196179280013137449136192463213005769 \epsilon \nonumber \\
    &+(64.58681327997710467911775172559510574\nonumber \\
    &-1.885876024599004396370902498544503624 i) \epsilon ^2\nonumber \\
    &-(482.8245437501750639199043360351808390\nonumber \\
    &-44.5252967502620937941700262685878594 i) \epsilon
   ^3\nonumber\\
   &+(2673.705654558192075753788726033738760\nonumber\\
   &-528.287832714112404395544526530310280 i) \epsilon
   ^4
\end{align}
\begin{align}\label{eq:num3}
    g_{34,\, E_1}^{\textbf{NPL1}} =& \frac{1}{2} -6.05565929110329624223788151979416509 \epsilon \nonumber\\
    &+31.7451124806499620748862588382041868 \epsilon
   ^2\nonumber\\
   &-107.8068744775645475416197961384826642 \epsilon^3\nonumber\\
   &+287.466453178744360269449569313085153 \epsilon ^4
\end{align}
\begin{align}\label{eq:num4}
    g_{34,\, P_1}^{\textbf{NPL1}} =& \frac{1}{2}-(13.1446725277151726799578082503121059356\nonumber\\
    &-4.7123889803846894949001937715189429169 i) \epsilon \nonumber\\
    &+(136.85472307581382279953083657017566781\nonumber\\
    &-109.85392819305858882573642117892036086 i) \epsilon
   ^2\nonumber\\
   &-(759.1389260208083012154864639749265974\nonumber\\
   &-1164.9603923525691646489930393582168744 i) \epsilon
   ^3\nonumber\\
   &+(2189.25480962727488495812790525445087\nonumber\\
   &-7667.89621488750185829424024333904334 i) \epsilon
   ^4
\end{align}
In appendix \ref{sec:ApA} we show explicit analytic results up to $\mathcal{O}(\epsilon^2)$ for all top-sector canonical master integrals.

In ancillary files attached to this publication we provide analytic results for all two-loop canonical master integrals in terms of MPLs. We provide files containing the solutions up to weight four. The weight five and six parts are too large to be included with the submitted files $(\sim 90$~\text{MB} for \textbf{PL1} and $\sim 240~\text{MB}$ for \textbf{NPL1}$)$. The interested reader can obtain them from~\cite{anc}. Additionally, we provide the canonical bases for families $\textbf{PL1}$ and $\textbf{NPL1}$, the corresponding differential equations, as well as closed-form solutions for all boundary terms. 

\section{Conclusion}\label{sec:Conclusion}
In this paper we present the analytic calculation of two-loop four-point master integrals for a planar and a non-planar topology with one massive leg and one massive propagator. These integrals are relevant to the non-factorisable corrections for t-channel single-top production at NNLO~\cite{Bronnum-Hansen:2021pqc}. 

We constructed canonical bases of master integrals~\cite{Henn:2013pwa} and used the Simplified Differential Equations approach~\cite{Papadopoulos:2014lla} to compute them analytically in terms of multiple polylogarithms~\cite{Goncharov:1998kja}. The necessary boundary terms were computed using the method of expansion-by-regions~\cite{Jantzen:2012mw} in closed-form in the dimensional regulator $\epsilon$. The structure of the canonical differential equations and the closed-form expressions of the boundary terms allowed us to obtain analytic solutions for all two-loop master integrals up to arbitrary order in $\epsilon$. 

We observe that going beyond $\mathcal{O}(\epsilon^4)$ results in highly cumbersome expressions, due to the appearance of huge numbers of multiple polylogarithms with transcendental weight five and six. We evaluated numerically~\cite{Duhr:2019tlz,Vollinga:2004sn} our expressions up to $\mathcal{O}(\epsilon^4)$, which involve at most weight four multiple polylogarithms, for Euclidean and physical phase-space points and cross-checked our results against \textsc{pySecDec}~\cite{Borowka:2017idc} and \textsc{AMFlow}~\cite{Liu:2022chg}.

Our results mark the first step in the analytic computation of all two-loop master integrals that are relevant to the non-factorisable corrections for t-channel single-top production at NNLO. We leave the calculation of the remaining integral families, as well as the optimisation of our analytic solutions for fast and efficient numerical evaluations in the physical region for future work.

\acknowledgments
The author would like to thank Dhimiter D. Canko for many fruitful discussions in various stages of this project, and Xiao Liu for help with the use of \textsc{AMFlow}. This work was supported by the Excellence Cluster ORIGINS funded by the Deutsche Forschungsgemeinschaft (DFG, German Research Foundation) under Germany's Excellence Strategy - EXC-2094 - 390783311.

\appendix

\section{Explicit results up to $\mathcal{O}(\epsilon^2)$}\label{sec:ApA}
We show explicit results up to $\mathcal{O}(\epsilon^2)$ for all top-sector canonical master integrals. We remind the reader that the square roots $r_1, r_2, r_3$ are functions of $\{S_{ij},m_w\}$ and are defined in \eqref{eq:roots}. 

\paragraph{Family $\textbf{PL1}$}
\begin{align}
    g_{29} =&\, \epsilon^2 \Bigg[-G\left(\frac{m_w^2}{S_{12}},\frac{r_1-S_{12}}{2 S_{12}},x\right)-G\left(\frac{m_w^2}{S_{12}},-\frac{r_1+S_{12}}{2 S_{12}},x\right)+2
   G\left(\frac{m_w^2}{S_{12}},0,x\right)\nonumber \\
   &-\log \left(m_w^2\right) G\left(\frac{m_w^2}{S_{12}},x\right)+\log \left(-S_{12}\right)
   G\left(\frac{m_w^2}{S_{12}},x\right)-G\left(-1,\frac{r_1-S_{12}}{2 S_{12}},x\right)\nonumber\\
   &-G\left(-1,-\frac{r_1+S_{12}}{2 S_{12}},x\right)+G\left(0,\frac{r_1-S_{12}}{2
   S_{12}},x\right)+G\left(0,-\frac{r_1+S_{12}}{2 S_{12}},x\right) \Bigg] + \mathcal{O}(\epsilon^3) 
\end{align}
\begin{align}
    g_{30} =&\,\frac{1}{4} + \epsilon \Bigg[-G\left(-\frac{m_w^2}{S_{23}},x\right)+\frac{1}{2} G\left(\frac{r_1-S_{12}}{2 S_{12}},x\right)+\frac{1}{2} G\left(-\frac{r_1+S_{12}}{2
   S_{12}},x\right)\nonumber \\
   &-G(0,x)-\frac{1}{2} \log \left(-S_{12}\right) \Bigg]
   +\epsilon^2 \Bigg[G\left(\frac{m_w^2}{S_{12}},\frac{r_1-S_{12}}{2 S_{12}},x\right)+G\left(\frac{m_w^2}{S_{12}},-\frac{r_1+S_{12}}{2 S_{12}},x\right)\nonumber \\
   &-\frac{1}{2} G\left(\frac{r_1-S_{12}}{2
   S_{12}},-\frac{m_w^2}{S_{23}},x\right)-\frac{1}{2} G\left(-\frac{r_1+S_{12}}{2 S_{12}},-\frac{m_w^2}{S_{23}},x\right)-2
   G\left(-\frac{m_w^2}{S_{23}},\frac{r_1-S_{12}}{2 S_{12}},x\right)\nonumber \\
   &-2 G\left(-\frac{m_w^2}{S_{23}},-\frac{r_1+S_{12}}{2 S_{12}},x\right)-\log \left(m_w^2\right)
   G\left(\frac{r_1-S_{12}}{2 S_{12}},x\right)-\log \left(m_w^2\right) G\left(-\frac{r_1+S_{12}}{2 S_{12}},x\right)\nonumber\\
   &+G\left(0,-\frac{m_w^2}{S_{23}},x\right)-2
   G\left(\frac{m_w^2}{S_{12}},0,x\right)+4 G\left(-\frac{m_w^2}{S_{23}},0,x\right)+4 G\left(-\frac{m_w^2}{S_{23}},-\frac{m_w^2}{S_{23}},x\right)\nonumber\\
   &-\frac{3}{2}
   G\left(-\frac{S_{12}+S_{23}}{S_{12}},-\frac{m_w^2}{S_{23}},x\right)-\log \left(-S_{12}\right) G\left(\frac{m_w^2}{S_{12}},x\right)+2 \log \left(-S_{12}\right)
   G\left(-\frac{m_w^2}{S_{23}},x\right)\nonumber\\
   &+\log \left(m_w^2\right) G\left(\frac{m_w^2}{S_{12}},x\right)-\frac{1}{2} G\left(0,\frac{r_1-S_{12}}{2
   S_{12}},x\right)-\frac{1}{2} G\left(0,-\frac{r_1+S_{12}}{2 S_{12}},x\right)\nonumber\\
   &-\frac{3}{2} G\left(\frac{r_1-S_{12}}{2 S_{12}},\frac{r_1-S_{12}}{2
   S_{12}},x\right)-\frac{3}{2} G\left(\frac{r_1-S_{12}}{2 S_{12}},-\frac{r_1+S_{12}}{2 S_{12}},x\right)\nonumber\\
   &-\frac{3}{2} G\left(-\frac{r_1+S_{12}}{2
   S_{12}},\frac{r_1-S_{12}}{2 S_{12}},x\right)-\frac{3}{2} G\left(-\frac{r_1+S_{12}}{2 S_{12}},-\frac{r_1+S_{12}}{2 S_{12}},x\right)\nonumber\\
   &+\frac{3}{2}
   G\left(-\frac{S_{12}+S_{23}}{S_{12}},\frac{r_1-S_{12}}{2 S_{12}},x\right)+\frac{3}{2} G\left(-\frac{S_{12}+S_{23}}{S_{12}},-\frac{r_1+S_{12}}{2 S_{12}},x\right)\nonumber\\
   &+2
   \log \left(-S_{12}\right) G(0,x)+4 G(0,0,x)+\frac{1}{2} \log ^2\left(-S_{12}\right)+\frac{\pi ^2}{24} \Bigg] + \mathcal{O}(\epsilon^3)
\end{align}
\begin{align}
    g_{31} =&\,-\frac{1}{4}+\epsilon \Bigg[G\left(-\frac{m_w^2}{S_{23}},x\right)-\frac{1}{2} G\left(\frac{r_1-S_{12}}{2 S_{12}},x\right)-\frac{1}{2} G\left(-\frac{r_1+S_{12}}{2 S_{12}},x\right)+G(0,x)\nonumber\\
    &+\frac{1}{2}
   \log \left(-S_{12}\right) \Bigg] +\epsilon^2 \Bigg[-G\left(\frac{m_w^2}{S_{12}},\frac{r_1-S_{12}}{2 S_{12}},x\right)-G\left(\frac{m_w^2}{S_{12}},-\frac{r_1+S_{12}}{2 S_{12}},x\right)\nonumber\\
   &+G\left(\frac{r_1-S_{12}}{2
   S_{12}},-\frac{m_w^2}{S_{23}},x\right)+G\left(-\frac{r_1+S_{12}}{2 S_{12}},-\frac{m_w^2}{S_{23}},x\right)+2 G\left(-\frac{m_w^2}{S_{23}},\frac{r_1-S_{12}}{2
   S_{12}},x\right)\nonumber\\
   &+2 G\left(-\frac{m_w^2}{S_{23}},-\frac{r_1+S_{12}}{2 S_{12}},x\right)+\log \left(m_w^2\right) G\left(\frac{r_1-S_{12}}{2 S_{12}},x\right)+\log
   \left(m_w^2\right) G\left(-\frac{r_1+S_{12}}{2 S_{12}},x\right)\nonumber\\
   &-2 G\left(0,-\frac{m_w^2}{S_{23}},x\right)+2 G\left(\frac{m_w^2}{S_{12}},0,x\right)-4
   G\left(-\frac{m_w^2}{S_{23}},0,x\right)-4
   G\left(-\frac{m_w^2}{S_{23}},-\frac{m_w^2}{S_{23}},x\right)\nonumber\\
   &+G\left(-\frac{S_{12}+S_{23}}{S_{12}},-\frac{m_w^2}{S_{23}},x\right)+\log \left(-S_{12}\right)
   G\left(\frac{m_w^2}{S_{12}},x\right)-2 \log \left(-S_{12}\right) G\left(-\frac{m_w^2}{S_{23}},x\right)\nonumber\\
   &-\log \left(m_w^2\right)
   G\left(\frac{m_w^2}{S_{12}},x\right)+G\left(0,\frac{r_1-S_{12}}{2 S_{12}},x\right)+G\left(0,-\frac{r_1+S_{12}}{2 S_{12}},x\right)\nonumber\\
   &+G\left(\frac{r_1-S_{12}}{2
   S_{12}},\frac{r_1-S_{12}}{2 S_{12}},x\right)+G\left(\frac{r_1-S_{12}}{2 S_{12}},-\frac{r_1+S_{12}}{2 S_{12}},x\right)\nonumber\\
   &+G\left(-\frac{r_1+S_{12}}{2
   S_{12}},\frac{r_1-S_{12}}{2 S_{12}},x\right)
   +G\left(-\frac{r_1+S_{12}}{2 S_{12}},-\frac{r_1+S_{12}}{2
   S_{12}},x\right)\nonumber\\
   &-G\left(-\frac{S_{12}+S_{23}}{S_{12}},\frac{r_1-S_{12}}{2 S_{12}},x\right)
   -G\left(-\frac{S_{12}+S_{23}}{S_{12}},-\frac{r_1+S_{12}}{2
   S_{12}},x\right)\nonumber\\
   &-2 \log \left(-S_{12}\right) G(0,x)-4 G(0,0,x)-\frac{1}{2} \log ^2\left(-S_{12}\right)-\frac{5 \pi ^2}{24} \Bigg]+\mathcal{O}(\epsilon^3)
\end{align}

\paragraph{Family $\textbf{NPL1}$}
\begin{align}
    g_{33} =&\, \frac{3}{4} +\epsilon \Bigg[-\frac{1}{2} G\left(-\frac{m_w^2}{S_{23}},x\right)-\frac{1}{2} G\left(\frac{m_w^2}{S_{12}+S_{23}},x\right)+G\left(\frac{r_1-S_{12}}{2
   S_{12}},x\right)\nonumber\\
   &+G\left(-\frac{r_1+S_{12}}{2 S_{12}},x\right)-3 G(0,x)-\frac{3}{2} \log \left(-S_{12}\right) \Bigg]
   +\epsilon^2 \Bigg[\frac{3}{2} \log ^2\left(-S_{12}\right)+6 G(0,x) \log \left(-S_{12}\right)\nonumber\\
   &-2 G\left(\frac{r_1-S_{12}}{2 S_{12}},x\right) \log \left(-S_{12}\right)-2
   G\left(-\frac{r_1+S_{12}}{2 S_{12}},x\right) \log \left(-S_{12}\right)+G\left(-\frac{m_w^2}{S_{23}},x\right) \log
   \left(-S_{12}\right)\nonumber\\
   &+G\left(\frac{m_w^2}{S_{12}+S_{23}},x\right) \log \left(-S_{12}\right)+12 G(0,0,x)+2 G\left(0,\frac{r_1-S_{12}}{2 S_{12}},x\right)+2
   G\left(0,-\frac{r_1+S_{12}}{2 S_{12}},x\right)\nonumber\\
   &-2 G\left(0,-\frac{m_w^2}{S_{23}},x\right)-2 G\left(0,\frac{m_w^2}{S_{12}+S_{23}},x\right)-4 G\left(\frac{r_1-S_{12}}{2
   S_{12}},0,x\right)\nonumber\\
   &-4 G\left(\frac{r_1-S_{12}}{2 S_{12}},\frac{r_1-S_{12}}{2 S_{12}},x\right)-4 G\left(\frac{r_1-S_{12}}{2 S_{12}},-\frac{r_1+S_{12}}{2
   S_{12}},x\right)+2 G\left(\frac{r_1-S_{12}}{2 S_{12}},-\frac{m_w^2}{S_{23}},x\right)\nonumber\\
   &+2 G\left(\frac{r_1-S_{12}}{2 S_{12}},\frac{m_w^2}{S_{12}+S_{23}},x\right)-4
   G\left(-\frac{r_1+S_{12}}{2 S_{12}},0,x\right)-4 G\left(-\frac{r_1+S_{12}}{2 S_{12}},\frac{r_1-S_{12}}{2 S_{12}},x\right)\nonumber\\
   &-4 G\left(-\frac{r_1+S_{12}}{2
   S_{12}},-\frac{r_1+S_{12}}{2 S_{12}},x\right)+2 G\left(-\frac{r_1+S_{12}}{2 S_{12}},-\frac{m_w^2}{S_{23}},x\right)\nonumber\\
   &+2 G\left(-\frac{r_1+S_{12}}{2
   S_{12}},\frac{m_w^2}{S_{12}+S_{23}},x\right)+2 G\left(-\frac{m_w^2}{S_{23}},0,x\right)-G\left(-\frac{m_w^2}{S_{23}},\frac{r_1-S_{12}}{2
   S_{12}},x\right)\nonumber\\
   &-G\left(-\frac{m_w^2}{S_{23}},-\frac{r_1+S_{12}}{2 S_{12}},x\right)+2 G\left(-\frac{m_w^2}{S_{23}},-\frac{m_w^2}{S_{23}},x\right)+2
   G\left(\frac{S_{23}}{S_{12}},\frac{r_1-S_{12}}{2 S_{12}},x\right)\nonumber\\
   &+2 G\left(\frac{S_{23}}{S_{12}},-\frac{r_1+S_{12}}{2 S_{12}},x\right)-2
   G\left(\frac{S_{23}}{S_{12}},\frac{m_w^2}{S_{12}+S_{23}},x\right)+2
   G\left(\frac{m_w^2}{S_{12}+S_{23}},0,x\right)\nonumber\\
   &-G\left(\frac{m_w^2}{S_{12}+S_{23}},\frac{r_1-S_{12}}{2
   S_{12}},x\right)-G\left(\frac{m_w^2}{S_{12}+S_{23}},-\frac{r_1+S_{12}}{2 S_{12}},x\right)\nonumber\\
   &+2 G\left(\frac{m_w^2}{S_{12}+S_{23}},\frac{m_w^2}{S_{12}+S_{23}},x\right)+2
   G\left(-\frac{S_{12}+S_{23}}{S_{12}},\frac{r_1-S_{12}}{2 S_{12}},x\right)\nonumber\\
   &+2 G\left(-\frac{S_{12}+S_{23}}{S_{12}},-\frac{r_1+S_{12}}{2 S_{12}},x\right)-2
   G\left(-\frac{S_{12}+S_{23}}{S_{12}},-\frac{m_w^2}{S_{23}},x\right)-\frac{23 \pi ^2}{24} \Bigg]+\mathcal{O}(\epsilon^3)
\end{align}
\begin{align}
   g_{34} =&\, \frac{1}{2} +\epsilon \Bigg[\frac{1}{2} G\left(-\frac{m_w^2}{S_{23}},x\right)-\frac{1}{2} G\left(\frac{m_w^2}{S_{12}+S_{23}},x\right)-\frac{1}{2} G\left(\frac{r_1-S_{12}}{2
   S_{12}},x\right)-\frac{1}{2} G\left(-\frac{r_1+S_{12}}{2 S_{12}},x\right)\nonumber\\
   &-2 G(0,x)-\log \left(-S_{12}\right) \Bigg]
   +\epsilon^2 \Bigg[\log ^2\left(-S_{12}\right)+4 G(0,x) \log \left(-S_{12}\right)\nonumber\\
   &+G\left(\frac{r_1-S_{12}}{2 S_{12}},x\right) \log \left(-S_{12}\right)+G\left(-\frac{r_1+S_{12}}{2
   S_{12}},x\right) \log \left(-S_{12}\right)-G\left(-\frac{m_w^2}{S_{23}},x\right) \log \left(-S_{12}\right)\nonumber\\
   &+G\left(\frac{m_w^2}{S_{12}+S_{23}},x\right) \log
   \left(-S_{12}\right)+8 G(0,0,x)-\frac{1}{2} G\left(0,\frac{r_1-S_{12}}{2 S_{12}},x\right)-\frac{1}{2} G\left(0,-\frac{r_1+S_{12}}{2 S_{12}},x\right)\nonumber\\
   &+\frac{1}{2}
   G\left(0,\frac{m_w^2}{S_{12}+S_{23}},x\right)+2 G\left(\frac{r_1-S_{12}}{2 S_{12}},0,x\right)+\frac{5}{2} G\left(\frac{r_1-S_{12}}{2 S_{12}},\frac{r_1-S_{12}}{2
   S_{12}},x\right)\nonumber\\
   &+\frac{5}{2} G\left(\frac{r_1-S_{12}}{2 S_{12}},-\frac{r_1+S_{12}}{2 S_{12}},x\right)-\frac{3}{2} G\left(\frac{r_1-S_{12}}{2
   S_{12}},-\frac{m_w^2}{S_{23}},x\right)-G\left(\frac{r_1-S_{12}}{2 S_{12}},\frac{m_w^2}{S_{12}+S_{23}},x\right)\nonumber\\
   &+2 G\left(-\frac{r_1+S_{12}}{2
   S_{12}},0,x\right)+\frac{5}{2} G\left(-\frac{r_1+S_{12}}{2 S_{12}},\frac{r_1-S_{12}}{2 S_{12}},x\right)+\frac{5}{2} G\left(-\frac{r_1+S_{12}}{2
   S_{12}},-\frac{r_1+S_{12}}{2 S_{12}},x\right)\nonumber\\
   &-\frac{3}{2} G\left(-\frac{r_1+S_{12}}{2 S_{12}},-\frac{m_w^2}{S_{23}},x\right)-G\left(-\frac{r_1+S_{12}}{2
   S_{12}},\frac{m_w^2}{S_{12}+S_{23}},x\right)-2 G\left(-\frac{m_w^2}{S_{23}},0,x\right)\nonumber\\
   &+G\left(-\frac{m_w^2}{S_{23}},\frac{r_1-S_{12}}{2
   S_{12}},x\right)+G\left(-\frac{m_w^2}{S_{23}},-\frac{r_1+S_{12}}{2 S_{12}},x\right)-2 G\left(-\frac{m_w^2}{S_{23}},-\frac{m_w^2}{S_{23}},x\right)\nonumber\\
   &+2
   G\left(\frac{m_w^2}{S_{12}+S_{23}},0,x\right)-G\left(\frac{m_w^2}{S_{12}+S_{23}},\frac{r_1-S_{12}}{2
   S_{12}},x\right)-G\left(\frac{m_w^2}{S_{12}+S_{23}},-\frac{r_1+S_{12}}{2 S_{12}},x\right)\nonumber\\
   &+2
   G\left(\frac{m_w^2}{S_{12}+S_{23}},\frac{m_w^2}{S_{12}+S_{23}},x\right)-\frac{5}{2} G\left(-\frac{S_{12}+S_{23}}{S_{12}},\frac{r_1-S_{12}}{2
   S_{12}},x\right)\nonumber\\
   &-\frac{5}{2} G\left(-\frac{S_{12}+S_{23}}{S_{12}},-\frac{r_1+S_{12}}{2 S_{12}},x\right)+\frac{5}{2}
   G\left(-\frac{S_{12}+S_{23}}{S_{12}},-\frac{m_w^2}{S_{23}},x\right)-\frac{5 \pi ^2}{12} \Bigg]+\mathcal{O}(\epsilon^3)
\end{align}
\begin{align}
    g_{35} =&\, \frac{1}{4} +\epsilon \Bigg[G\left(\frac{m_w^2}{S_{12}+S_{23}},x\right)-G(0,x)-\frac{1}{2} \log \left(-S_{12}\right) \Bigg] \nonumber\\
    &+\epsilon^2 \Bigg[2 G\left(\frac{m_w^2}{S_{12}+S_{23}},\frac{r_1-S_{12}}{2 S_{12}},x\right)+2 G\left(\frac{m_w^2}{S_{12}+S_{23}},-\frac{r_1+S_{12}}{2 S_{12}},x\right)+2
   G\left(0,-\frac{m_w^2}{S_{23}},x\right)\nonumber\\
   &+2 G\left(0,\frac{m_w^2}{S_{12}+S_{23}},x\right)+G\left(\frac{S_{23}}{S_{12}},\frac{m_w^2}{S_{12}+S_{23}},x\right)-4
   G\left(\frac{m_w^2}{S_{12}+S_{23}},0,x\right)\nonumber\\
   &-4
   G\left(\frac{m_w^2}{S_{12}+S_{23}},\frac{m_w^2}{S_{12}+S_{23}},x\right)-G\left(-\frac{S_{12}+S_{23}}{S_{12}},-\frac{m_w^2}{S_{23}},x\right)\nonumber\\
   &-2 \log
   \left(-S_{12}\right) G\left(\frac{m_w^2}{S_{12}+S_{23}},x\right)
   -2 G\left(0,\frac{r_1-S_{12}}{2 S_{12}},x\right)-2 G\left(0,-\frac{r_1+S_{12}}{2
   S_{12}},x\right)\nonumber\\
   &-G\left(\frac{S_{23}}{S_{12}},\frac{r_1-S_{12}}{2 S_{12}},x\right)-G\left(\frac{S_{23}}{S_{12}},-\frac{r_1+S_{12}}{2
   S_{12}},x\right)+G\left(-\frac{S_{12}+S_{23}}{S_{12}},\frac{r_1-S_{12}}{2 S_{12}},x\right)\nonumber\\
   &+G\left(-\frac{S_{12}+S_{23}}{S_{12}},-\frac{r_1+S_{12}}{2
   S_{12}},x\right)+2 \log \left(-S_{12}\right) G(0,x)+4 G(0,0,x)+\frac{1}{2} \log ^2\left(-S_{12}\right)-\frac{\pi ^2}{24} \Bigg]\nonumber\\
   &+\mathcal{O}(\epsilon^3)
\end{align}



\bibliographystyle{JHEP}

\bibliography{main.bib}

\end{document}